\documentclass[%
 reprint,
 superscriptaddress,
 amsmath,amssymb,
 aps,
 prb,
]{revtex4-2}

\usepackage[dvipsnames]{xcolor}
\usepackage{physics}
\usepackage{quantikz}
\usepackage{float}
\usepackage{graphicx}
\usepackage{dcolumn}
\usepackage{bm}
\usepackage{tikz}
\usepackage[T1]{fontenc}
\usetikzlibrary{matrix, decorations.pathreplacing, positioning}
\usepackage{float}

\definecolor{C1}{RGB}{52, 89, 149}
\definecolor{C2}{RGB}{251, 77, 61}
\definecolor{C3}{RGB}{3, 206, 164}
\definecolor{C4}{RGB}{202, 21, 81}
\usepackage{hyperref}
\hypersetup{colorlinks=true, linkcolor=C2, citecolor=C2, urlcolor=C2}

\usepackage[normalem]{ulem}
\usepackage{lipsum}

\newlength\figureheight
\newlength\figurewidth

\renewcommand{\selectlanguage}[1]{}

\begin{document}

\title{Quantum quenches of scar states in the Affleck-Kennedy-Lieb-Tasaki model\\via Clifford augmented tensor network simulation}

\author{Azar C.\ Nakhl}
\email{azar_nakhl@sutd.edu.sg}
\affiliation{Science Mathematics and Technology Cluster, Singapore University of Technology and Design, 8 Somapah Road, 487372 Singapore}
\affiliation{School of Physics, The University of Melbourne, Parkville, Victoria 3010, Australia}
\author{Ben Harper}
\affiliation{School of Physics, The University of Melbourne, Parkville, Victoria 3010, Australia}
\affiliation{Data61, CSIRO, Clayton, Victoria 3168, Australia}
\author{Muhammad Usman}
\affiliation{Faculty of Information Technology, Monash University, Clayton, Victoria 3800, Australia}
\affiliation{School of Physics, The University of Melbourne, Parkville, Victoria 3010, Australia}
\affiliation{Data61, CSIRO, Clayton, Victoria 3168, Australia}
\author{Thomas Quella}
\affiliation{School of Mathematics and Statistics, The University of Melbourne, Parkville, Victoria 3010, Australia}

\date{\today}
\begin{abstract}
  Clifford augmented methods have emerged as a powerful technique to simulate quantum circuits and many-body systems by exploiting the stabiliser structure found in said systems, whilst additionally enabling one to readily quantify the non-stabilizerness (i.e.\ magic) present. In this work, we utilise a qudit Clifford augmented simulation method to probe the presence of magic in the Affleck-Kennedy-Lieb-Tasaki (AKLT) model, finding that its ground state and scar states are highly magical as quantified by their Stabiliser Rényi Entropy (SRE). Nevertheless, owing to its low entanglement structure, these states are still readily expressible as a Clifford augmented Matrix Product State with low bond-dimension and moreover may be time-evolved using the Clifford augmented time-dependent variational principle (TDVP) algorithm. We utilise Clifford augmented TDVP to study the string order dynamics in the scar states of the AKLT model after a quench within the Haldane phase to the anti-ferromagnetic Heisenberg Hamiltonian. We observe that there is a decay of string order following the quench, which is more pronounced with increasing bimagnon number. This finding highlights the sensitivity of string order in the AKLT scar states to symmetry-preserving quenches that break the underlying restricted spectrum generating algebra, even when the initial and final Hamiltonians belong to the same topological phase.
\end{abstract}

\maketitle

\section{Introduction}
The simulation of many-body quantum systems at scale continues to be a computationally challenging task, with tensor network methods serving as the primary way to find ground states and time evolve systems which exhibit low entanglement~\cite{vidal_efficient_2003,PhysRevLett.93.040502,haegeman2011time,haegeman2014unifying}. With the introduction of Clifford augmented tensor network methods~\cite{qian_augmenting_2024, lami_quantum_2024, masot-llima_stabilizer_2024,nakhl_stabilizer_2025,harper_gcamps_2025}, the limiting resource shifts from entanglement to the non-stabilizerness (or magic) present within the system. Notably, the Clifford augmented Density Matrix Renormalisation Group (DMRG) algorithm~\cite{schollwock2011density} for finding ground-states of spin-systems~\cite{qian_augmenting_2024} has been found useful in extracting greater simulation accuracy given some fixed bond-dimension as demonstrated through simulations of the $J_1-J_2$ Heisenberg model~\cite{qian_augmenting_2024}, the critical Ising and XXZ spin chains~\cite{fan_disentangling_2025}, and \textit{ab-initio} quantum chemistry calculations~\cite{fu2025clifford}.

From a quantum resource theory perspective, there has also been considerable interest in the study of non-stabilizerness in spin systems~\cite{viscardi_interplay_2025,rattacaso_stabilizer_2023,odavic_stabilizer_2025,hoshino_stabilizer_2025,frau_stabilizer_2025,nehra2025topological,tarabunga_many-body_2023,physrevlett.133.010601,oliviero_magic-state_2022,tarabunga2024critical,jasser2025stabilizer} where it has primarily served as an indicator of different phenomena in these systems, namely criticality~\cite{viscardi_interplay_2025,oliviero_magic-state_2022,haug2023quantifying,tarabunga_many-body_2023,physrevb.110.045101} and symmetry-protected topological (SPT) phases~\cite{nehra2025topological}. More recently, there has been an effort to adapt algorithms for time evolution~\cite{qian_clifford_2025, mello_clifford_2025} and finite-temperature simulations~\cite{qian_augmenting_2024-1}. The former of these facilitated simulations of the dynamics of the XXZ model and Kitaev spin liquid~\cite{qian_clifford_2025}, Loschmidt echos of non-integrable Ising chains~\cite{mello2025clifford}, as well as flux tube dynamics in a 2+1D $\mathbb{Z}_2$ Higgs gauge theory \cite{Bacciconi:2026arXiv260827561B}.

Despite the ongoing research into the non-stabilizerness of spin-1 chains and qudit systems more broadly~\cite{physrevb.110.045101,magni_quantum_2025}, the lack of a general qudit simulation framework has, for a long time, inhibited algorithms such as Clifford augmented DMRG and time-dependent variation principle (TDVP) algorithm from being used in these higher-spin settings. This only changed recently with the introduction of GCAMPS in Ref.~\cite{harper_gcamps_2025}. GCAMPS enables us to consider spin-$1$ systems using fewer computational resources than required from conventional ground state such as DMRG~\cite{schollwock2011density}, time evolution algorithms such as the Time-Evolved Block Decimation algorithm (TEBD)~\cite{PhysRevLett.93.040502} and TDVP~\cite{,haegeman2011time,haegeman2014unifying}. In this work we use the GCAMPS algorithm to study two aspects of scar states \cite{Moudgalya:2018PhRvB..98w5155M,moudgalya2018entanglement,mark2020unified} in the Affleck-Kennedy-Lieb-Tasaki (AKLT) model~\cite{affleck1987rigorous}: their non-stabilizerness properties and the stability of their topological features under a quench within the Haldane phase.

The AKLT model is an isotropic spin-1 chain that was historically proposed \cite{affleck1987rigorous} as a mathematically tractable realization of a Haldane phase \cite{Haldane:PhysRevLett.50.1153}, i.e.\ a phase with a gap above a unique ground state. The ground state has an exact representation as a Matrix Product State (MPS), exhibits non-trivial string order \cite{DenNijs:PhysRevB.40.4709} and was later identified to reside in a non-trivial SPT phase with respect to any of the following symmetries: $SO(3)$ rotation symmetry, the subgroup $\mathbb{Z}_2\times\mathbb{Z}_2$ of $\pi$-rotations about the principal axes, time-reversal and bond-centered inversion \cite{Pollmann:PhysRevB.81.064439,Pollmann:2012PhRvB..85g5125P,Chen:PhysRevB.83.035107,Schuch:1010.3732v3}. The model is also known to admit towers of exact excited eigenstates \cite{Moudgalya:2018PhRvB..98w5155M,moudgalya2018entanglement,mark2020unified}, so-called scar states, that are arising from certain bimagnon excitations and exhibit surprising physical features such as atypically low entanglement, violating the usual volume-law entanglement of thermal states \cite{Turner:2018NatPh..14..745T}.

More recently, it was argued that suitable subspaces of scar states collectively exhibit topological properties that mimic those of the ground state \cite{matsui2025symmetry}. In individual scar states that leads specifically to the presence of non-trivial string order. In Ref.\ \cite{matsui2025symmetry} it was stressed that these features are intimately linked to the presence of a protecting symmetry, in conjunction with a restricted Spectrum Generating Algebra (rSGA) \cite{mark2020unified,Moudgalya:2020PhRvB.102h5140M}. However, it remained unclear whether the topological properties of scar subspaces are stable under generic deformations away from the AKLT Hamiltonian which break up the subspace structure while staying in the symmetry-protected Haldane phase. In this work, we study this question through the numerical analysis of quenched dynamics whose simulation has been enabled by GCAMPS.

More specifically, we compute the time evolution of exact scar eigenstates of the AKLT Hamiltonian and their string order after a quench to the anti-ferromagnetic Heisenberg Hamiltonian. While this quench takes place entirely {\em within} the Haldane phase \cite{Nightingale:1986PhRvB..33..659N,white1993numerical} and all relevant symmetries protecting the SPT phase are preserved, the algebraic structure supporting the scar subspaces, the rSGA, breaks down. Our work complements studies where scar states have been investigated from the viewpoint of thermalization after perturbation \cite{Lin:2020PhRvR...2c3044L,Sanada:2023PhRvB.108o5102S} or where the fate of string order has been investigated for the ground state \cite{Mazza:2014PhRvB..90b0301M,calvanese2016destruction,Hagymasi:2019PhRvL.122y0601H,Dhar:2022PhRvB.105i4309D}, especially when quenching in or out of the Haldane phase.

As our numerical tool, GCAMPS, expresses qudit quantum states as Clifford augmented MPS, our approach is tailored to physical situations that feature both non-trivial entanglement and non-stabilizerness. We observe that the AKLT ground state and scar excitations exhibit high non-stabilizerness as measured by the Stabilizer Rényi Entropy (SRE), but maintain relatively low Non-stabilizerness Entanglement Entropy (NsEE)~\cite{huang_non-stabilizerness_2024}, consistent with the low entanglement entropy exhibited by these states.

After the quench to the Heisenberg model, we find that the Clifford augmented TDVP exhibits a higher simulation accuracy compared to a regular MPS TDVP algorithm with the same maximum bond dimension, in line with earlier investigations~\cite{qian_clifford_2025,mello_clifford_2025}. However, while these previous works time evolved from either a stabilizer state~\cite{mello_clifford_2025,qian_clifford_2025} or state with low non-stabilizerness~\cite{qian_clifford_2025}, our simulations have an initial state with high non-stabilizerness. Yet we continue to see an advantage in the Clifford augmented protocol. This enables us to evolve the system over a longer timescale whilst keeping the error within an acceptable range enabling us to probe the string order of the scar states following this deformation.

Our simulations reveal a dependence of the string order dynamics on the bimagnon number, with a more rapid decay for the higher-energy scar states studied. This behaviour shows that the string order inherited from the AKLT ground state through the rSGA responds differently across the scar tower to a quench that removes the supporting rSGA while preserving the Hamiltonian's topological protecting symmetries. The results provide insight into the robustness of the string order beyond the AKLT ground state. Whether the observed decay ultimately leads to vanishing long-range string order remains an open question requiring simulations over a longer time-scale. The tensor networks that would facilitate such simulations necessitate a high maximum bond dimension requiring significant memory and very high runtime costs, which we leave for future studies.

The remainder of this work is organised as follows: In Section~\ref{sec:methods} we give a brief overview of the Clifford augmented TDVP algorithm used and the measures of magic which we compute. In Section~\ref{sec:bbsc} we introduce the bilinear biquadratic spin chain, including the AKLT model and its scar subspaces. We then describe the quench to the antiferromagnetic Heisenberg Hamiltonian which we will study. In Section~\ref{sec:simres} we initially investigate the non-stabilizerness of the AKLT ground state and scar states before presenting results of the quench for the ground state and the first three scar states. Lastly, in Section~\ref{sec:conc} we summarise our findings and provide some possible future directions. 
\begin{figure}
    \centering
    \includegraphics[width=\linewidth]{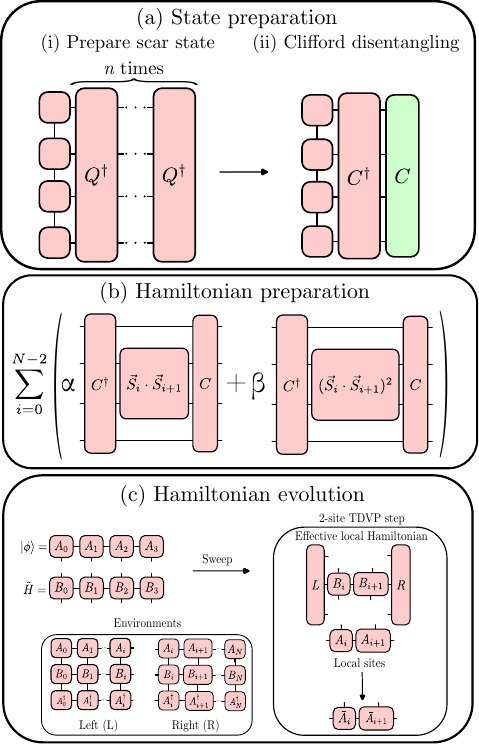}
    \caption{An overview of the Clifford augmented simulation method: (a) One prepares the scar states by repeated application of $Q^\dagger$ defined in Equation~\eqref{eq:q} after which one may disentangle the scar state $\ket{S_i}$ using the Clifford operator $C^\dagger$ (green). (b) To time-evolve $\ket{S_i} = C\ket{\text{MPS}}$ with $H$ one prepares the Clifford augmented Hamiltonian $\tilde{H} = C^\dagger H C$. As $H$ is composed of nearest neighbour spin operators one can build $\tilde{H}$ directly given the stabilizer tableau representation of $C$ and the decompositions of $S^i$ in Appendix~\ref{appen:tdvp}. (c) An overview of the 2-site TDVP procedure for the Clifford augmented MPS $\ket{\psi}$ and Hamiltonian $\tilde{H}$, this algorithm is unchanged from the typical TDVP algorithm for MPS.}
    \label{fig:fig_1}
\end{figure}
\section{Methods} \label{sec:methods}
In this section, we will give a brief overview of the setup enabling the time evolution simulations conducted in this work and the method used to find the SRE of the states. Precise details are given in Appendix~\ref{appen:tdvp} for the Clifford augmented TDVP algorithm and in Appendix~\ref{appen:sre} for the SRE computations. In Figure~\ref{fig:fig_1} we provide a high-level overview of preparing a Clifford augmented state and how one time evolves said state using TDVP. 

\subsection{Tensor network time evolution algorithms}
There exists a number of MPS algorithms for time-evolution of a quantum system, including the Time Evolved Block Decimation (TEBD) algorithm~\cite{PhysRevLett.93.040502}, the time-dependent Density Matrix Renormalisation Group (tDMRG) algorithm~\cite{daley2004time}, and the Time Dependent Variational Principle (TDVP) algorithm~\cite{haegeman2011time,haegeman2014unifying}. Of these, TDVP is the most applicable in a Clifford augmented setting as the evolution is applied locally regardless of the locality of the Hamiltonian. Both single-site~\cite{mello_clifford_2025} and two-site~\cite{qian_clifford_2025} Clifford-augmented TDVP algorithms have been proposed for the evolution of systems with low initial non-stabilizerness. In this work we use the two-site Clifford augmented TDVP method. Further details about the 2-site TDVP, including the translation of spin-1 Hamiltonians to the qutrit Clifford basis can be found in Appendix~\ref{appen:tdvp}.

\subsection{Computing measures of magic}
As we aim to investigate the non-stabilizerness of the quantum spin chains we simulate one must first consider which measures of non-stabilizerness may be determined efficiently. We will focus on two such measures, firstly the Stabilizer Rényi Entropy (SRE)~\cite{leone_stabilizer_2022} which is a faithful magic monotone for $\alpha\geq2$ as well as the non-stabilizerness entanglement entropy (NsEE)~\cite{huang_non-stabilizerness_2024} which is well approximated by states represented with GCAMPS. We give brief definitions of these measures below, with further details found in Appendix~\ref{appen:sre}.
\subsubsection{Stabilizer Rényi Entropy}
SRE is defined by
\begin{equation}
   M_\alpha(\ket{\psi}) = \frac{1}{1-\alpha}\log_d \left(\sum_{P\in\mathcal{P}_N} \frac{\bra{\psi}P\ket{\psi}^{2\alpha}}{d^{N\alpha}}\right) - N, \label{eq:sre}
\end{equation}
where $\alpha$ is the Rényi index, $\mathcal{P}_N$ is the Pauli group over $N$ sites, and $d$ is the dimension of the local Hilbert space. Later we will specialise to $d=3$ for the spin-1 systems under consideration. $M_\alpha$ as defined in Equation~\eqref{eq:sre} is a measure of non-stabilizerness, which unlike minimal $T$-gate count does not require an ideal circuit decomposition. Provided that one may represent the state using an MPS with low bond dimension, the computation of SRE may be performed more efficiently by first converting the state to a \textit{Pauli MPS}, which is a representation of the state in terms of the Pauli basis. In Appendix~\ref{appen:sre} we detail how one constructs the Pauli MPS of a state represented in GCAMPS and how to compute the SRE in this representation.
\subsubsection{Non-Stabilizerness Entanglement Entropy}
In the context of (G)CAMPS in particular the NsEE serves as a computationally relatively inexpensive way of determining non-stabilizerness provided that the disentangling routine is performed in a globally optimal fashion and that one does not require a faithful magic monotone (i.e.\ $M(\ket{\psi}) = 0 \iff \ket{\psi} \text{ is a stabilizer state}$). The NsEE is found by simply considering the entanglement entropy of the tensor network part within (G)CAMPS. We will present results including an approximate NsEE in Section~\ref{sec:simres}.

\section{The Bilinear Biquadratic Spin Chain} \label{sec:bbsc}
The spin systems we consider in this work are special instances of the spin-$1$ bilinear biquadratic spin chain whose Hamiltonian is given by
\begin{equation}
    H(\theta) = J\sum_{i=0}^{N-1} \left(\cos(\theta) \vec{S}_i\cdot \vec{S}_{i+1} + \sin(\theta) (\vec{S}_i\cdot \vec{S}_{i+1})^2\right). \label{eq:bilin}
\end{equation}
We consider Periodic Boundary Conditions (PBC), hence index ``$N$'' corresponds to site $0$. The $\vec{S}$ is a vector of the spin-$1$ operators given in Equation~\eqref{eq:spin}. The system has been extensively studied as its ground states admit a number of interesting quantum phases found in different regions of $\theta$ \cite{Lauchli:2006PhRvB..74n4426L}. In this work we will be focusing specifically on systems within the \textit{Haldane phase} which is situated in the parameter range $-\frac{\pi}{4}<\theta<\frac{\pi}{4}$. The Haldane phase is characterised by a finite energy gap between its ground and excited states \cite{affleck1987rigorous} and known to correspond to a Symmetry Protected Topological (SPT) phase \cite{Pollmann:PhysRevB.81.064439,Pollmann:2012PhRvB..85g5125P,Chen:PhysRevB.83.035107,Schuch:1010.3732v3}. It exhibits a range of properties, including non-trivial string order \cite{DenNijs:PhysRevB.40.4709}, that are robust against perturbations that preserve certain symmetries such as the $SO(3)$ invariance inherent in the Hamiltonian \eqref{eq:bilin}, as long as the gap does not close.

\subsection{The Affleck–Kennedy–Lieb–Tasaki model}

  A well-studied point in the Haldane phase is the AKLT model \cite{affleck1987rigorous} whose Hamiltonian is given by 
\begin{align}
    H_{\text{AKLT}} &= \sum_{i=0}^{N-1} \left(\frac{1}{2}\vec{S}_i\cdot \vec{S}_{i+1} + \frac{1}{6} (\vec{S}_i\cdot\vec{S}_{i+1})^2 + \frac{1}{3}I\right). \label{eq:aklt}
\end{align}
It arises from the Hamiltonian in Equation~\eqref{eq:bilin} by choosing $J=\frac{\sqrt{10}}{6}$ and $\theta=\arctan\left(\frac{1}{3}\right)$. There is also an extra identity term added so that the ground state satisfies $\expval{H_{\text{AKLT}}}=0$. The AKLT model is notable as its ground state $\ket{S_0}$ is exactly solvable~\cite{affleck1987rigorous}, having a compact MPS representation of bond dimension $2$. Given the general mathematical expression for an MPS 
\begin{equation}
    \ket{\psi} = \sum_{i_0,i_1,\dots,i_{N-1}} \text{Tr}(A_{i_0}^{(0)}A_{i_1}^{(1)}\dots A_{i_{N-1}}^{(N-1)} )\ket{i_0i_1\dots i_{N-1}}, \label{eq:mathmps}
\end{equation}
the $A^{(i)}_{j_i}$ of the ground state are translationally invariant (i.e.\ do not vary with $i$), and are given by
\begin{align}
    A_1 &= \sqrt{\frac{2}{3}}\sigma^+=\sqrt{\frac{2}{3}}\begin{bmatrix}
        0 &1\\0&0
    \end{bmatrix} \\
    A_0&=-\sqrt{\frac{1}{3}}Z = -\sqrt{\frac{1}{3}}\begin{bmatrix}
        1 & 0\\0&-1
    \end{bmatrix}\\
    A_{-1}&=-\sqrt{\frac{2}{3}}\sigma^-=-\sqrt{\frac{2}{3}}\begin{bmatrix}
        0 & 0 \\1&0
    \end{bmatrix}
\end{align}
corresponding to the three $S^z$ eigenvalues $1,0,-1$. As a brief aside, the OBC AKLT model in which the sum in Equation~\eqref{eq:aklt} terminates at $N-2$, similarly admits an MPS representation with bond dimension $2$, with the edge tensors determined by projecting onto one of the spin-$1/2$ edge states, i.e.
\begin{equation}
    A^{(0)}_{i_0} = v_l A_{i_0}, \quad A^{(N-1)}_{i_{N-1}} = A_{i_{N-1}}v_r 
\end{equation}
for vectors $v_l = \begin{bmatrix}
   1 & 0 
\end{bmatrix} \text{ or } \begin{bmatrix}
    0 & 1 
\end{bmatrix}$ and $v_r=\begin{bmatrix}
    1\\0
\end{bmatrix} \text{ or } \begin{bmatrix}
    0\\1
\end{bmatrix}$, and with the $A_{i_j}$ given as before. The tensors in the bulk remain unchanged from the periodic case. The OBC ground state is $4$-fold degenerate for the four possible combinations of $v_l$ and $v_r$. 

We represent the Hamiltonian in Equations~\eqref{eq:bilin} and~\eqref{eq:aklt} using a Matrix Product Operator (MPO) which we build up from the identity $\text{MPO} =I_3\otimes I_3\otimes \dots\otimes I_3$ by applying each of the nearest neighbour spin operators onto the MPO. The MPO that encodes the AKLT Hamiltonian has a bond dimension of $10$ in the bulk. 

The simulation framework we use based off Ref.~\cite{harper_gcamps_2025} is limited to OBC MPS, hence we embed the PBC ground state and Hamiltonians into MPS and MPO with OBC. Note that Clifford optimisation should (ideally) preserve the translational invariance of the tensors in the bulk of the MPS. Implementing the PBC AKLT Hamiltonian with an OBC MPO results in an increase of the bond dimension in the bulk to $18$. For the PBC ground state embedded into an OBC MPS the bond dimension in the bulk is $4$. 

In addition to the ground state, the AKLT model admits a number of exact low-entanglement scars~\cite{moudgalya2018entanglement} which may be built atop the ground state as
\begin{align}
   \ket{S_{2n}} &= (Q^\dagger)^n\ket{\psi_0}, \\
   Q&=\sum_{i=0}^{N-1} (-1)^i(S^+_i)^2, \label{eq:q} \\
   S^+&=\sqrt{2}\begin{bmatrix}
   0&1&0 \\
   0&0&1 \\
   0&0&0
   \end{bmatrix}
\end{align}
for bimagnon number $n=0,1,\dots,N/2$ and AKLT ground state $\ket{\psi_0}$. Note that if $N/2$ is odd then $\ket{S_{N}}=0$ which means that in this case one does not arrive at the ferromagnetic $\ket{11\dots 1}$ state~\cite{mark2020unified}. In this work we will assume even values for $N$. The energy of the scar states is $\bra{S_{2n}}H_{\text{AKLT}}\ket{S_{2n}}=2n$ which is independent of chain length~\cite{moudgalya2018entanglement}. Let us also emphasize that the power $(Q^\dagger)^n$ of the bimagnon operator has a compact representation as an MPO \cite{moudgalya2018entanglement} which in turn implies an MPS representation of scar states, independent of system size, and hence their atypically low entanglement \cite{mark2020unified}. There exist other scar towers that are related to the previous one by the action of symmetries. Collectively, these scar towers played a key role in the recent discussion of topological properties that can be attributed to scar subspaces \cite{matsui2025symmetry}.

A metric that is relevant to our simulations is the string order parameter~\cite{DenNijs:PhysRevB.40.4709}, which is used as a signature of SPT phases. It is a measure of long-range correlation given by
\begin{equation}
    O^\alpha_\text{string}(l_1,l_2) = \bra{\psi}S^\alpha_{i_1} \exp(i\pi\sum_{i_1<j<i_2}S^\alpha_j) S^\alpha_{i_2}\ket{\psi}.
\end{equation}
In the translationally invariant PBC case which we will be focusing on, we can express the string order simply in terms of the distance $r$ as,
\begin{equation}
    O^\alpha_\text{string}(r) = \bra{\psi}S^\alpha_{0} \exp(i\pi\sum_{0<j<r}S^\alpha_j) S^\alpha_{r}\ket{\psi}. \label{eq:so}
\end{equation}
For the AKLT ground state the string order is $-\frac{4}{9}$ in the thermodynamic limit~\cite{PhysRevB.45.304,kennedy1992hidden} (i.e.\ for $N\rightarrow \infty$, followed by $r\rightarrow\infty$). Recently it was found that the scar states share a finite long distance string order in the thermodynamic limit~\cite{matsui2025symmetry}, providing one of several signatures that the scar subspace may inherit the topological properties of the ground state through the presence of the rSGA.

\subsection{Quenching within the Haldane Phase}
We will be performing a quench, i.e.\ abruptly changing the Hamiltonian under which the state unitarily evolves, starting from the ground state and low energy scar states of the AKLT model introduced above and evolving under the anti-ferromagnetic Heisenberg model given by Equation~\eqref{eq:bilin} with $\theta=0$, that is,
\begin{equation}
    H_H = J\sum_{i=0}^{N-1} \vec{S}_i\cdot \vec{S}_{i+1}. \label{eq:heis}
\end{equation}
In performing this quench we seek to assess the stability of the ground and scar states and their string order.

Unlike the ground state and the excited scar states of the AKLT model, the spin-$1$ anti-ferromagnetic Heisenberg model has no known exactly solvable eigenstates. Numerical investigations established that it is still residing in the Haldane phase~\cite{white1993numerical}, with a long-range string order of $-0.374$ and ground state energy $\bra{\psi_H}{H_H}\ket{\psi_H} \approx-1.4N J$, where $\ket{\psi_H}$ is the ground state of the Heisenberg model and $N$ is the number of sites. Note that for the AKLT ground state $\bra{S_0}{H_H}\ket{S_0}=-\frac{4N}{3}J$.

As it will become relevant to the numerics, we note that the quench should conserve the Heisenberg energy $\bra{\psi(t)}{H_H}\ket{\psi(t)}$ where $\ket{\psi(t)}$ is the time-evolved state. This is because the time-evolution operator $e^{i H_H t}$ commutes with $H_H$ giving rise to the identity
\begin{equation}
     \bra{\psi(0)}e^{-i H_H t}H_He^{i H_H t}\ket{\psi(0)}=\bra{\psi(0)}H_H\ket{\psi(0)}
\end{equation}
for the Heisenberg energy of the time-evolved state, where $\ket{\psi(0)}$ is the initial state. Numerically though, this does not necessarily hold for a $2$-site TDVP after a certain time if the time step $\Delta t$ is too large or the maximum bond dimension is too small. Hence we will use the deviation in $\expval{H_H}$ given  by $\epsilon(t)=|\bra{\psi(0)}H_H\ket{\psi(0)}-\bra{\psi(t)}H_H\ket{\psi(t)}|/N$ as a measure for the error accumulated in the simulation.

\section{Results and Discussion}\label{sec:simres}
In the following section we will initially assess the non-stabilizerness in the AKLT ground and scar states, before considering a quench of the AKLT ground state and low-energy scar states to the anti-ferromagnetic Heisenberg Hamiltonian using a 2-site TDVP algorithm on both GCAMPS and conventional MPS. Throughout we will be considering a system of $N=50$ sites with PBC.

\subsection{Non-stabilizerness vs entanglement in the AKLT model}
Firstly, comparing entanglement and non-stabilizerness, we note that the ground state is characterised by a translationally invariant MPS of bond dimension $2$, with the entanglement entropy at a constant $S=2\log_2(2)=2$ for all partitions of size $N/2$. For the scar states, for fixed bimagnon number $n$, the entanglement entropy also satisfies an area law, whereas for $n\sim N$ one finds a sub-volume law entanglement entropy scaling in the number of sites $N$~\cite{moudgalya2018entanglement}, i.e.\ scaling as $S\sim\log_2(N)$. 

\begin{figure}
    \centering
    \includegraphics[width=\linewidth]{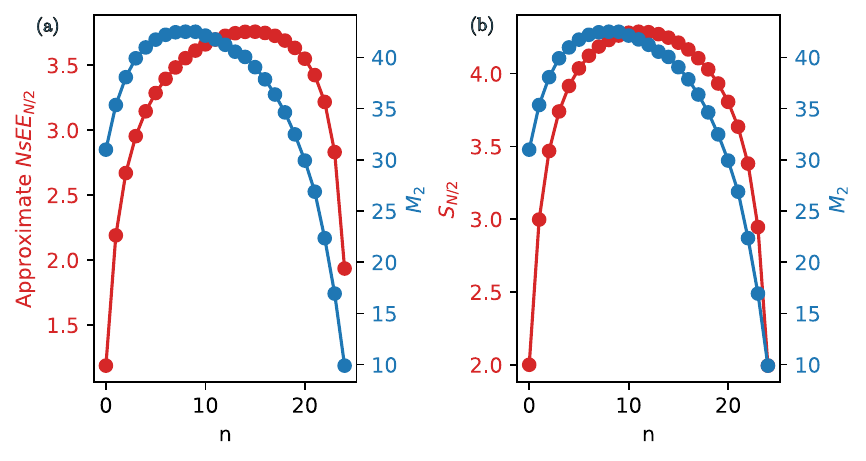}
    \caption[Entanglement and Magic of the AKLT scar states]{(a) A plot of the approximate NsEE (in red) and SRE (in blue) and (b) a plot of the Entanglement entropy (in red) and SRE (in blue) for the AKLT ground state and tower of scar states for $N=50$ sites. The bimagnon number $n$ labels each of the scar states, with the ground state given by $n=0$. We find that SRE peaks first at about $n=9$ with entanglement entropy peaking at the already established $n=N/4$~\cite{mark2020unified}. The approximate NsEE continues to grow until $n=15$, indicating sub-optimal disentangling with GCAMPS. Note that the approximate NsEE serves as a measure of computational complexity and is distinct from the true NsEE which like SRE grows with increasing non-stabilizerness.}
    \label{fig:sre_nsee_ent}
\end{figure}

We numerically find the SRE and NsEE for the AKLT and the tower of scar states and plot it in Figure~\ref{fig:sre_nsee_ent}. Note that the NsEE is approximate as the disentangling performed is only over a linear chain of sites, repeating until convergence in the entanglement entropy of the MPS is reached. This is not necessarily the optimal disentangling needed to precisely characterise the NsEE. Indeed we find that compared to both SRE and entanglement entropy, the NsEE continues to increase after $n>N/4$ where both other metrics have peaked. This demonstrates that the approximate NsEE presented is not a magic monotone, but is illustrative of the computational complexity of the GCAMPS simulations performed. We would still like to stress that the approximate NsEE remains lower than the entanglement entropy for any given $n$.

We find the SRE for each of the scar states by sweeping across a number of approximate GCAMPS MPS with increasing maximum bond dimension, terminating the procedure when the SRE value converges to within $10^{-2}$ of the previously determined value. We note that using an approximate GCAMPS MPS performs as well as using an approximate Pauli MPS with the former not requiring one to determine the full Pauli MPS with bond dimension $\chi_P=\chi^2$ at any point which is particularly advantageous for the relatively highly-entangled scar states mid-spectrum.

Interestingly we find that the SRE of the AKLT ground state and many of the lower energy scar states is quite high. Indeed, we observe that the SRE of the scar states peaks slightly before $n=N/4$ with close to maximal SRE, noting that for a $D$-dimensional Hilbert space an upper bound for the SRE is given by~\cite{cuffaro2024quantum}
\begin{equation}
    M_\alpha(\ket{\psi}) \leq \frac{1}{1-\alpha}\log_d(\frac{1+(D-1)(D+1)^{1-\alpha}}{D}).
\end{equation}
For our $50$ site systems then the 2-SRE with $D=3^{50}$ is bounded by $M_2\lesssim 49.33$. Note that this upper bound is not necessarily tight, with a tighter upper bound having been established for two-qubit states~\cite{liu2026maximal}. As the highest energy scar state is the ferromagnetic $\ket{111\dots1}$ state, the SRE eventually drops to $0$, however for odd $N/2$ $\ket{S_{N}}=\vec{0}$~\cite{mark2020unified}, hence this is not present in Figure~\ref{fig:sre_nsee_ent} which goes up to $n=N/2 -1$.  Finally, we note that the SRE appears to peak earlier than the entanglement entropy, which is a consequence of the high SRE at $n=0$ and that SRE is $0$ at $n=N/2$ and the finite growth in magic possible by application of the $Q^\dagger$ operator, noting that the increase in magic from the application of any operator is bounded. 

\subsection{Evolution of string order and entanglement entropy} \label{sec:so}
\begin{figure}
    \centering
    \includegraphics[width=1\linewidth]{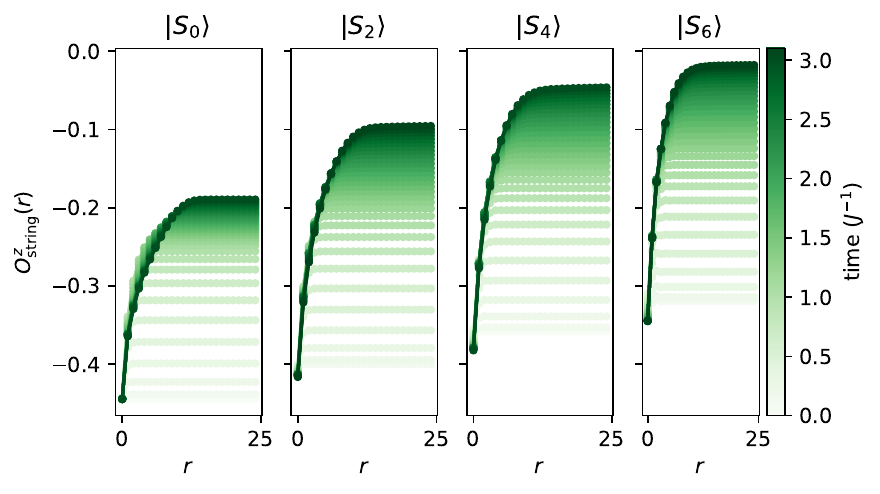}
    \caption[String order after a quantum quench in the Haldane phase]{Plot of the string order $O^z_\text{string}(r)$ as per Equation~\eqref{eq:so} after a quench to the isotropic Heisenberg Hamiltonian starting from the ground state $\ket{S_0}$ and first three scar states ($\ket{S_2}$, $\ket{S_4}$, $\ket{S_6}$) of the AKLT model with $50$ sites. The time-step of the TDVP simulation was set to $\Delta t =0.001 J^{-1}$ and plotted at every $t=0.1 J^{-1}$. The bond dimension of the MPS set to a maximum of $\chi=512$ with the disentangling performed once at the beginning of the time evolution. Note that the AKLT ground state has string order $O^z_\text{string}(r)\approx-4/9$ for all string lengths $r$ and that the isotropic Heisenberg Hamiltonian has long range string order $O^z_\text{string}(N/2)\approx -0.374$~\cite{white1993numerical}. Additionally, note that for each $r$ we take an average over all contiguous blocks of that size as translational invariance is (ideally) maintained by this quench.}
    \label{fig:string_order}
\end{figure}

We now consider the case of a quantum quench of the AKLT ground state $\ket{S_0}$ (i.e.\ the $0$\textsuperscript{th} scar state) and first three scar states, $\ket{S_2}$, $\ket{S_4}$ and $\ket{S_6}$ respectively, to time evolution under the anti-ferromagnetic Heisenberg Hamiltonian. We simulate this system for chain lengths $30-60$, with time-step $\Delta t = 0.001~J^{-1}$ and maximum bond-dimension $\chi = 512$. We present the results of the string order for the $50$ site case in Figure~\ref{fig:string_order}. Note as the TDVP algorithm does not inherently preserve the translational invariance of the system when one limits the growth in bond-dimension, we average the string order for some given distance $r$ over all possible pairs of sites $|l_2-l_1| = r$. Indeed, we recognise that the breaking of translational invariance could be used as a qualitative measure of simulation accuracy.

For the ground state, we largely find that the string order is independent of the string length $r$ with its absolute value decreasing with $t$ when $t\lesssim 1 J^{-1}$. This qualitative behaviour changes for later times ($t\gtrsim 2 J^{-1}$) though with the string order decaying with increasing $r$ until approximately $N/4$. This effect appears to be more pronounced with the scar states which indeed do not appear to precisely plateau to a fixed value of the string order parameter at large string lengths.

\begin{figure*}
    \centering
    \includegraphics[scale=1]{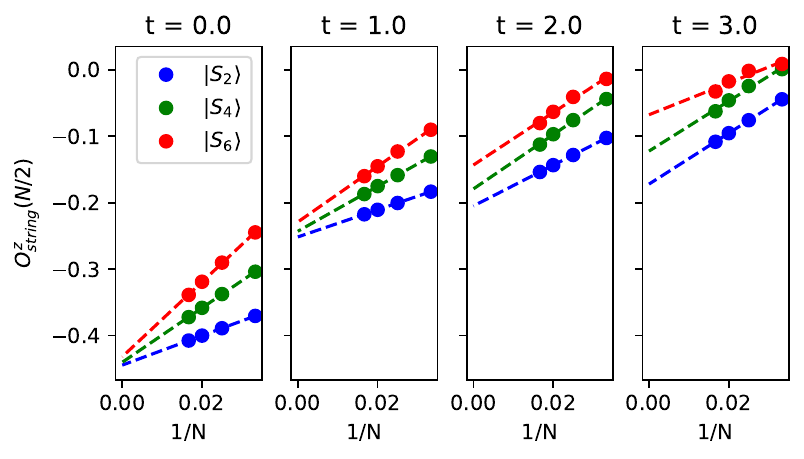}
    \caption[Long distance string order for the first three scar states]{A $1/N$ fit (dashed lines) for $O^z_{string}(N/2)$ for the $\ket{S_2}$, $\ket{S_4}$ and $\ket{S_6}$ scars taken at regular time intervals after quenching as described in Section~\ref{sec:so}. After some time we find that the fit breaks down for the higher-energy scars as the string order decays to zero.}
    \label{fig:string_order_Fit}
\end{figure*}
\begin{figure}
    \centering
    \includegraphics[width=\linewidth]{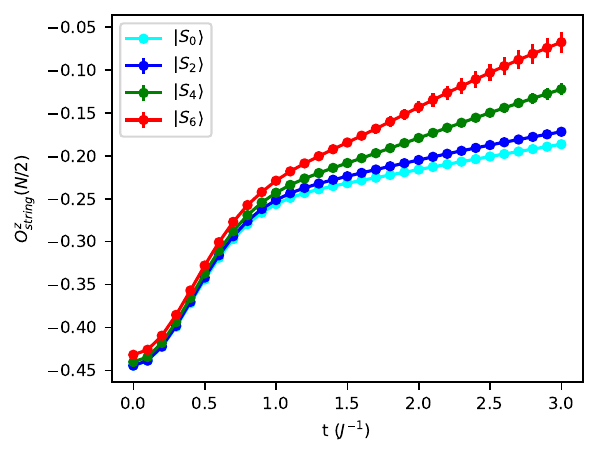}
    \caption[Long distance order parameter during a quantum quench]{A plot of $O^z_{string}(N/2)$ in the long length limit as estimated from Figure~\ref{fig:string_order_Fit} with the error bars showing one standard deviation in the fitted parameters. Note that the string order appears to decay to zero for higher-energy scar states after a short time, with the ground state and first scar state exhibiting comparatively slower decay over the same time interval.}
    \label{fig:long_range_so}
\end{figure}

Next we find the string order in the long distance and thermodynamic (i.e.\ $N\to\infty$) limits, noting that the scar states have a long distance string order that scales as $1/N$~\cite{quella_revision}. We plot this fit for various times in Figure~\ref{fig:string_order_Fit}. We find that at early times this fit works reasonably well, but appears to break down somewhat at later times for the higher energy scars as the string order decays to zero. Applying this procedure across all time steps gives us the evolution of the long distance string order in the thermodynamic limit for $0\leq t\leq3$ as shown in Figure~\ref{fig:long_range_so}. Although not precisely converged over the time range simulated here, it appears that the string order ``melts''~\cite{Mazza:2014PhRvB..90b0301M,calvanese2016destruction,Kairys:2022PhRvR...4d3189K} (i.e.\ rapidly decays to zero) for the $\ket{S_4}$ and $\ket{S_6}$ scars. We interpret this as an indication of the thermalisation of the state, however a full analysis of this would require a simulation over a longer time-scale with a much larger bond-dimension to analyse the volume-law entanglement-growth of the state. The ground state and $\ket{S_2}$ scar state on the other hand appear to have a long-range string order converging to $\approx -0.23$ and $\approx -0.21$ respectively which is less ordered than that of the Heisenberg long-range string order of $-0.374$~\cite{white1993numerical}. We note that a decay of string order is not necessarily surprising but rather to be expected. As stressed in Ref.\ \cite{matsui2025symmetry}, the string order of scar states in the AKLT model arises in connection with the topological properties tied to a suitable collection of scar towers associated with the rSGA, and the latter is absent in the Heisenberg model. It is also still worth noting that vanishing string order in itself is not necessarily an indication of the absence of topological properties \cite{Hagymasi:2019PhRvB..99g5145H}.
\begin{figure}
    \centering
    \includegraphics[width=\linewidth]{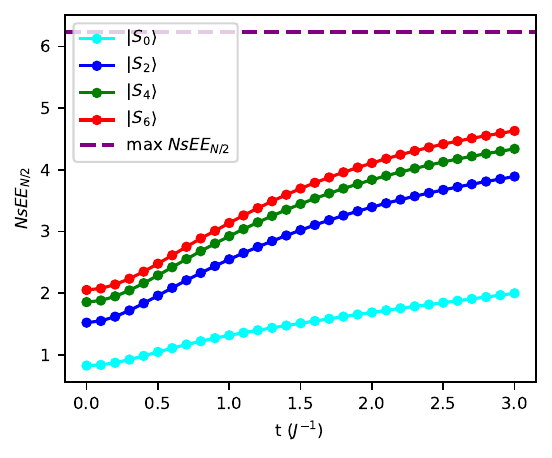}
    \caption[Evolution of Entanglement and non-stabilizerness during a quantum quench]{Half-chain approximate NsEE for the $50$ site quenches shown in Figure~\ref{fig:string_order}. The maximum possible entropy is given by $\log(512)\approx6.24$ (shown by the dashed line). The entanglement entropy for states represented in GCAMPS requires a reconstruction of the state as an MPS which is not performed here. We note that the NsEE appears to plateau for the scar states well below the maximum possible entropy, with only a small increase in NsEE with increasing $n$. For smaller chain lengths, we see a finite-size effect in which there is a sharp increase in entropy once the maximum bond dimension is reached.}
    \label{fig:ent_evol}
\end{figure}

Lastly, we consider the half-chain approximate NsEE of the quench in Figure~\ref{fig:string_order}. We again note that due to the lack of translational invariance this quantity is sensitive to the ordering of the sites on the MPS, but unlike the case for string order it is not feasible to consider an average over all possible $N/2$ sites. We find that the NsEE gradually increases with time for all initial states, and appears to increase with the energy of the initial state as well (i.e.\ with $n$). Noting that the maximum entropy is given by $S=\ln(\chi)\approx6.24$ we find that the states do not reach the maximum possible NsEE theoretically accessible, we however should stress that this is not necessarily an indication that giving the state access to a greater Hilbert space would not result in an increase in NsEE. To determine the ``regular'' entanglement entropy one would need to apply the disentangling Clifford $C$ to the MPS to reconstruct the state entirely in the tensor network, however this is a costly operation which would result in an increase in the bond dimension of the MPS and is hence not performed here.
\begin{figure}
    \centering
    \includegraphics[width=0.95\linewidth]{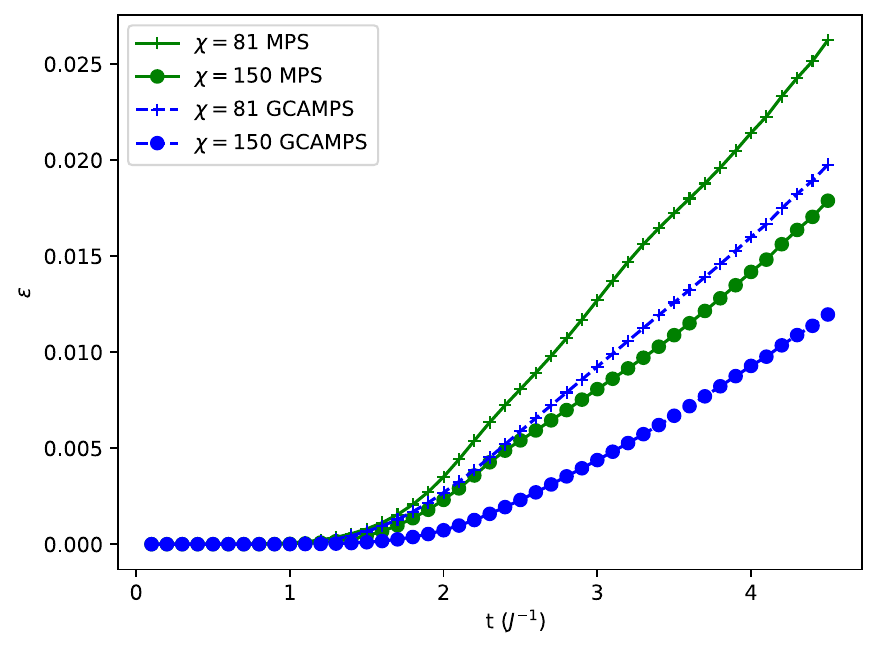}
    \caption[Error in Heisenberg energy with GCAMPS and MPS]{Plot of the deviation $\epsilon = \frac{1}{N}|\expval{H_H}(0) - \expval{H_H}(t)|$ in the Heisenberg energy for both GCAMPS and standard MPS simulation at different bond dimensions. The simulation parameters for the GCAMPS data were similar to that of Figure~\ref{fig:string_order} but with a larger TDVP time-step of $\Delta t =0.01 J^{-1}$. We observe that for GCAMPS simulation the error $\epsilon$ is similar to that of an MPS with approximately twice the bond dimension.}
    \label{fig:gcamps_mps_err}
\end{figure}
\begin{figure}
    \includegraphics[width=\linewidth]{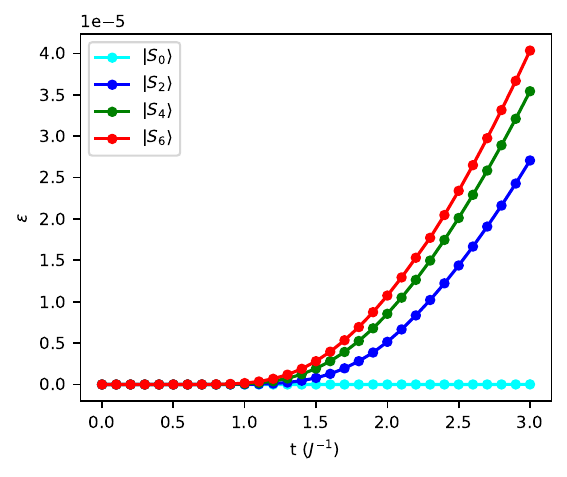}
    \caption[Error in Heisenberg Energy for the AKLT scar states]{Deviation in the Heisenberg energy $\epsilon = \frac{1}{N}|\expval{H_H}(0) - \expval{H_H}(t)|$ for GCAMPS of the $60$ site quench for the ground state and first three scar states. We find that the error accumulated for the scar states appear to be similar, compared to the ground state which maintains a low error rate throughout.}
    \label{fig:err_in_evol}
\end{figure}
\subsection{Simulation Accuracy compared to MPS}
As one does not have access to the full Hilbert space for large system sizes nor is it possible to simulate a time-evolution with an infinitely small time-step, there will inherently be some accumulation of error as one time-evolves within some restricted Hilbert space and with a discrete time-step. Although Clifford augmentation does not do much to counteract a large time-step, it does grant access to regions of the full $d^N$-dimensional Hilbert space that are otherwise inaccessible to an MPS with fixed maximum bond dimension $\chi$. 

To assess the accuracy of the Clifford augmented TDVP compared to the conventional MPS TDVP algorithm, we quench the ground state of the AKLT model as before but with a time-step of $\Delta t =0.01 J^{-1}$ instead of the previously considered $\Delta t =0.001 J^{-1}$ and with maximum bond dimensions $\chi=81$ and $\chi=150$. The results are shown in Figure~\ref{fig:gcamps_mps_err}. In general it appears that the error in the Clifford augmented TDVP is roughly similar to that of the conventional MPS TDVP with approximately twice the maximum bond-dimension. This appears consistent with our hypothesis that Clifford augmentation enables access to a larger Hilbert space than that otherwise accessible with just an MPS with the same bond dimension.  

Additionally, we keep track of the errors in the quenches considered in Section~\ref{sec:so} for both the ground state and first three scar states with results as shown in Figure~\ref{fig:err_in_evol}. We find that there is a significantly larger error and an earlier onset of deviations for the scar states compared to the ground state. We also note that the scar states appear to accumulate error at a similar rate to each other. Overall though the error rate shown in Figure~\ref{fig:err_in_evol} is well within an acceptable range throughout the evolution. We should stress that a negligible deviation in Heisenberg energy is only one possible error metric and does not necessarily guarantee that the state simulated is indeed the desired state. As previously mentioned, the preservation of translational invariance is another metric one could keep track of as it is not guaranteed to be preserved under the GCAMPS TDVP method used. 

\section{Conclusion} \label{sec:conc}

Building upon the sustained interest in studying non-stabilizerness in quantum many-body systems~\cite{viscardi_interplay_2025,rattacaso_stabilizer_2023,odavic_stabilizer_2025,hoshino_stabilizer_2025,frau_stabilizer_2025,nehra2025topological,tarabunga_many-body_2023,physrevlett.133.010601,oliviero_magic-state_2022,tarabunga2024critical,jasser2025stabilizer,fux_disentangling_2025,qian_augmenting_2024,qian_augmenting_2024-1,mello_clifford_2025} in this work we investigated the non-stabilizerness of the Affleck-Kennedy-Lieb-Tasaki (AKLT) model and assessed the ability of the GCAMPS simulator which we developed to simulate the dynamics of a quantum quench of this model. We found that, unlike the case for entanglement, the non-stabilizerness of the AKLT ground state and low-energy scar states is relatively high as measured by the Stabilizer Rényi Entropy (SRE). In contrast, the approximate Non-stabilizerness Entanglement Entropy (NsEE), an unfaithful measure of magic that is upper bounded by the entanglement entropy and serves as a direct measure of the computational complexity of GCAMPS simulation remained well below the entanglement entropy for the ground state and low-energy scar states. 

Studying the dynamics of a quench of the AKLT ground state and low-energy scar states to the anti-ferromagnetic Heisenberg Hamiltonian, we concluded that the GCAMPS simulation outperforms the same time-evolution algorithm on a regular MPS as measured by the deviation in the Heisenberg energy during the quench. Observing the string order during the quench we found evidence of string-order melting for higher-energy scar states. This observation is in line with the fact that the presence of string order is tied to the symmetry-protected topological properties of the whole scar subspace as opposed to individual scar states \cite{matsui2025symmetry} in the absence of a protecting rSGA. And this scar subspace structure breaks down in the absence of the rSGA at the Heisenberg point. However, based on the analysis of the NsEE and energy deviation at longer time scales, higher bond dimensions are required to completely capture the dynamics of this system and corroborate this finding further.

Beyond the aforementioned larger bond-dimension simulations, future directions could focus on incorporating Clifford augmentation to infinite MPS (iMPS)~\cite{PhysRevLett.98.070201} and variational uniform MPS (VUMPS)~\cite{PhysRevB.97.045145} allowing one to study the non-stabilizerness and perform dynamics of translationally invariant systems in a way that ensures preservation of the translational invariance. One may also wish to study quenches across topological phase transitions in the bilinear biquadratic spin model where one is more likely to observe a change in the non-stabilizerness dynamics, congruent with the spin-$1/2$ case~\cite{nehra2025topological}. Finally, for computation of SRE during the quench where the system is characterised using an MPS with a significantly larger bond dimension, sampling methods~\cite{physrevb.110.045101} may prove to be more computationally efficient without much loss in accuracy compared to the approximate GCAMPS method that was utilised in this work.
\bigskip

\subsection*{Note added in proof}

After this paper was submitted to the arXive we became aware of the work \cite{Ebner:2026arXiv261000129E} which also observed high non-stabilizerness in quantum many-body scars, specifically in 1+1D gauge theories.
\bigskip

\section*{Acknowledgements}
We would like to thank Chihiro Matsui for carefully reading the first version of this manuscript and providing feedback that helped shape the current version. This research was supported by the Commonwealth through an Australian Government Research Training Program Scholarship [DOI: \url{https://doi.org/10.82133/C42F-K220}], the University of Melbourne’s Research Computing Services, the Petascale Campus Initiative and the IBM Quantum Network Hub at the University of Melbourne. \\
\bibliography{merged,newref,bibliographyTQ,bib_ACN_new}
\cleardoublepage
\onecolumngrid

\appendix
\section{GCAMPS TDVP} \label{appen:tdvp}
Following~\cite{mendl_pytenet_2018, qian_clifford_2025} we perform a two-site Time Dependent Variational Principle (TDVP) simulation with a Clifford optimisation subroutine occurring at some interval $k$. In the Clifford augmented TDVP algorithm, one performs the time evolution entirely on the MPS, which necessitates commuting the time-evolution operator through a Clifford operator. For a state in the (G)CAMPS representation $\ket{\psi} = C\ket{\text{MPS}}$ where $C$ is a Clifford operator stored in a Stabilizer tableau, the state after a TDVP step $e^{iH\Delta t}\ket{\psi}$ is then equivalent to $e^{i\tilde{H}\Delta t}\ket{\text{MPS}}$ where $\tilde{H} = C^\dagger H C$. In the spin-$1/2$ case where Hamiltonians are typically written in terms of Pauli strings this conversion may be performed efficiently by multiplying rows of the (de)stabilizer tableau for $C$ which correspond to each of the $(X)Z$ which appear in the Pauli strings that make up the Hamiltonian. In the spin-$1$ case however, Hamiltonians are typically given in terms of the spin-$1$ operators which are given by,
\begin{align}
    S^x &= \frac{1}{\sqrt{2}}\begin{bmatrix}
        0 & 1&0\\1&0&1\\0&1&0
    \end{bmatrix}, \\ S^y&=\frac{1}{\sqrt{2}i}\begin{bmatrix}
        0&1&0\\
        -1&0&1\\0&-1&0
    \end{bmatrix},\\
    S^z &=\begin{bmatrix}
        1 &0&0\\0&0&0\\0&0&-1
    \end{bmatrix}. \label{eq:spin}
\end{align}
These are not the qutrit Pauli matrices, hence one must decompose these operators in terms of the qutrit Paulis. The decomposition, which we compute numerically is given by,
\begin{align}
    S^x &= \frac{\sqrt{2}}{3} X + \frac{\sqrt{2}}{3} X^2 - \frac{\sqrt{2}}{6}X^2 Z^2 - \frac{\sqrt{2}}{6} X^2 Z + \frac{\sqrt{2}}{6} e^{-i\pi/3} XZ + \frac{\sqrt{2}}{6}e^{i\pi/3}X Z^2 \\
    S^y &= \frac{i\sqrt{2}}{3} X - \frac{i\sqrt{2}}{3} X^2 + \frac{i\sqrt{2}}{6}X^2 Z^2 + \frac{i\sqrt{2}}{6}X^2 Z + \frac{i\sqrt{2}}{6} e^{-i\pi/3} XZ + \frac{i\sqrt{2}}{6}e^{i\pi/3}X Z^2 \\
    S^z &= \frac{1}{\sqrt{3}} e^{-i\pi/6} Z  + \frac{1}{\sqrt{3}} e^{i\pi/6} Z^2.
\end{align}
Constructing the Hamiltonian $H$ explicitly in terms of the above decompositions allows one to then construct $\tilde{H}$ using the same prescription as that for the spin-$1/2$ Hamiltonians. Following this, the TDVP algorithm proceeds as it would for a standard MPS representation with $\tilde{H}$ as the Hamiltonian and $\ket{\text{MPS}}$ as the state. We use the \texttt{pytenet}~\cite{mendl_pytenet_2018} package for performing the TDVP algorithm on the augmented Hamiltonian and MPS. After $k$ timesteps one performs the (G)CAMPS Clifford disentangling algorithm and reconstructs $\tilde{H}$ given the new Clifford operator before proceeding with the TDVP algorithm. The frequency with which one performs this disentangling is dependent on the system being simulated and whether one finds a reduction of the entanglement entropy of $\ket{\text{MPS}}$ by performing the disentangling algorithm.  

\section{The string order parameter in the Pauli basis}
For the purposes of computing the string order parameter $O^z_\text{string}(l_1,l_2)$ as per Equation~\eqref{eq:so} for a state in the GCAMPS representation, it is convenient to represent the operator $e^{i\pi S^z}$ in the qutrit Pauli basis. In said basis the operator takes the form,
\begin{equation}
    e^{i\pi S^z} =  \frac{-1}{3} I - \frac{2}{3}e^{i \pi/3} Z - \frac{2}{3}e^{-i \pi/3} Z^2. 
\end{equation}
One may then proceed to construct the operator as above and apply it directly to the MPS to find the string order parameter.
\section{Finding SRE using Pauli MPS} \label{appen:sre}
Given a state $\ket{\psi}$ in a physical system with $N$ sites and $d$ levels per site, the representation in the Pauli basis is given by
\begin{equation}
    \ketbra{\psi}{\psi}=\frac{1}{d^N}\sum_{i=0}^{d^{2N}-1} \bra{\psi}P_i\ket{\psi} P_i,
\end{equation}
where $i=i_0i_1\dots i_{N-1}$ is a $d^2$-nary number for the $N$-site Pauli string $P_i = P_{i_0}\otimes P_{i_1}\otimes \dots\otimes P_{i_{N-1}}$. The coefficients $\bra{\psi}P_i\ket{\psi}$ can be used to form a vector $\ket{P(\psi)}$ with elements $\expval{i|P(\psi)}=\bra{\psi}P_i\ket{\psi}/{\sqrt{d^N}}$. Given an MPS representation of $\ket{\psi}$ as per Equation~\eqref{eq:mathmps},
\begin{equation}
       \ket{\psi} = \sum_{i_0,i_1,\dots,i_{N-1}} A_{i_0}^{(0)}A_{i_1}^{(1)}\dots A_{i_{N-1}}^{(N-1)} \ket{i_0i_1\dots i_{N-1}},
\end{equation}
one can find the Pauli vector elements as
\begin{align}
    \bra{\psi}P_i\ket{\psi}/{\sqrt{d^N}} = \frac{1}{\sqrt{d^N}}\sum_{\substack{j_0,j_1,\dots,j_{N-1} \\ k_0,k_1,\dots, k_{N-1}}} &(A_{j_0}^{(0)\dagger}A_{j_1}^{(1)\dagger}\dots A_{j_{N-1}}^{(N-1)\dagger})(A_{k_0}^{(0)}A_{k_1}^{(1)}\dots A_{k_{N-1}}^{(N-1)})  \nonumber
    \\&\bra{j_0j_1\dots j_{N-1}} P_i \ket{k_0k_1\dots k_{N-1}}. \label{eq:pmps}
\end{align}
Noting that $\bra{j_0j_1\dots j_{N-1}}P_{i}\ket{k_0k_1\dots k_{N-1}} = \prod_{n=0}^{N-1} \bra{j_n}P_{i_n}\ket{k_n}$, we hence have that
\begin{align}
    \eqref{eq:pmps} &= \frac{1}{\sqrt{d^N}}\sum_{\substack{j_0,j_1,\dots,j_{N-1} \\ k_0,k_1,\dots,k_{N-1}}} \prod_{n=0}^{N-1} (A_{j_n}^{(n)\dagger} \otimes A_{k_n}^{(n)}) \bra{j_n}P_{i_n}\ket{k_n} \nonumber\\
    &= \frac{1}{\sqrt{d^N}} \prod_{n=0}^{N-1} B_{i_n}^{(n)},
\end{align}
where $B^{(n)}_{i_n} = \sum_{j=0}^{d-1}\sum_{k=0}^{d-1} \bra{j}P_{i_n} \ket{k} A^{(n)\dagger}_{j} \otimes A^{(n)}_{k}/\sqrt{d}$. Hence the Pauli MPS has the form 
\begin{equation}
    \ket{P(\psi)} = \sum_{i_0,i_1,\dots,i_{N-1}} B_{i_0}^{(0)}B_{i_1}^{(1)}\dots B_{i_{N-1}}^{(N-1)} \ket{i_0i_1\dots i_{N-1}}.
\end{equation}
Note the $B^{(i)}_{s_i}$ are of size $\chi_l^2 \times \chi_r^2$ (where the original matrix $A^{(i)}_{s_i}$ is of size $\chi_l \times \chi_r$). Graphically, to build the Pauli MPS, given the tensor at some site, the Pauli MPS tensor at that same site may be constructed as
\begin{equation*}
    \begin{array}{c}\includegraphics[]{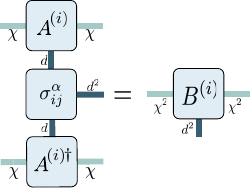}\end{array},
\end{equation*}
where $\sigma^\alpha_{ij}$ is a tensor containing all the single qudit Pauli operators (e.g.\ $\sigma^0_{ij}=I$, $\sigma^1_{ij}=X$, $\sigma^2_{ij}=Y$, $\sigma^3_{ij}=Z$ for the qubit case.). Performing this contraction over all sites gives one the Pauli MPS for the state.

To find the SRE, one needs to compute the quantity $\bra{\psi}P\ket{\psi}^{2\alpha}$ which given an MPS representation of the state can be framed in terms of finding the norm of a \textit{replica MPS} as introduced in Ref.~\cite{haug2023quantifying}. In the Pauli MPS formalism, the replica MPS is constructed using the MPO $W$ whose elements are given by $\bra{i}W\ket{i'} = \delta_{i,i'} \expval{i|P(\psi)}$, i.e.\  it is a diagonal operator whose elements correspond to the Pauli MPS elements. As a result the MPO of $W$ has the same bond dimension at each site as the Pauli MPS. Applying $W$ to the Pauli MPS $n-1$ times gives a state with the elements,
\begin{align}
    \bra{i}W^{n-1} \ket{P(\psi)} &= \bra{i}W(\sum_{i'} \ketbra{i'}{i'})W^{n-2}\ket{P(\psi)}\nonumber \\
    &= \frac{1}{\sqrt{d^N}}\sum_{i'} \delta_{i,i'} (\expval{i|P(\psi)}) \bra{i'} W^{n-2} \ket{P(\psi)}\nonumber \\
    &= \frac{1}{\sqrt{d^{N}}} \bra{\psi} P_{i} \ket{\psi} \bra{i} W^{n-2} \ket{P(\psi)} \nonumber\\
    &\qquad\qquad\qquad\vdots  \nonumber\\
    &= \frac{1}{\sqrt{d^{Nn}}} (\bra{\psi} P_{i} \ket{\psi})^{n}.
\end{align}
Taking the overlap of this with itself then gives us,
\begin{equation}
    \bra{P(\psi)} W^{(n-1)\dagger} W^{n-1} \ket{P(\psi)}  = \frac{1}{d^{Nn}}\sum_i (\bra{\psi} P_{i} \ket{\psi})^{2n},
\end{equation}
which is precisely what one needs to determine to find the SRE,
\begin{equation}
   M_\alpha(\ket{\psi}) = \frac{1}{1-\alpha} \log_d \left(\sum_{P\in\mathcal{P}_N} \frac{\bra{\psi}P\ket{\psi}^{2\alpha}}{d^{N\alpha}}\right) - N.
\end{equation}
Hence to find SRE one can first construct the Pauli MPS and MPO $W$, applying it $n-1$ times to the Pauli MPS and then compute the overlap which can all be done efficiently provided one has a low bond dimension MPS representation of the state $\ket{\psi}$. 

Finally, in the context of GCAMPS where $\ket{\psi}=C\ket{\text{MPS}}$ we note that the SRE is given by,
\begin{align}
   M_n(\ket{\psi}) &= \frac{1}{1-n} \log_d \left(\sum_{P\in\mathcal{P}_N} \frac{1}{d^ {Nn}}\bra{\text{MPS}}C^\dagger PC\ket{\text{MPS}}^{2n}\right) - N \\
   &= \frac{1}{1-n} \log_d \left(\sum_{\tilde{P}\in\mathcal{P}_N} \frac{1}{d^ {Nn}}\bra{\text{MPS}}\tilde{P}\ket{\text{MPS}}^{2n}\right) - N,
\end{align}
where $\tilde{P} = C^\dagger P C$ and noting that Clifford conjugation is a one-to-one mapping from one Pauli string to another. This means that one need only compute the Pauli MPS of $\ket{\text{MPS}}$ to find the SRE of $\ket{\psi}$ which is particularly useful in cases where the maximum bond dimension of $\ket{\text{MPS}}$ is smaller than that of $\ket{\psi}$.

\end{document}